# Experimental Demonstration of Energy Chirp Compensation by a Tunable Dielectric Based Structure

S. Antipov[1,3], S. Baturin[5], C. Jing[1,3], M. Fedurin[2], A. Kanareykin[1,5], C. Swinson[2], P. Schoessow[1], W. Gai[3] and A. Zholents[4].

[1]. Euclid Techlabs LLC, Solon, OH 44139

[2]. Accelerator Test Facility, Brookhaven National Laboratory, Upton, NY 11973

[3]. High Energy Physics Division, Argonne National Laboratory, Lemont, IL 60439

[4]. Advanced Photon Source, Argonne National Laboratory, Lemont, IL 60439

[5]. St.Petersburg Electrotechnical University LETI, St.Petersburg Russia 197376

**Abstract:**

A 60 MeV beam at the BNL Accelerator Test Facility (ATF) was manipulated by a planar tunable de-chirper made out of two 10 cm long dielectric slabs with copper plated backs. While the gap was changed from 5.8 mm to 1 mm, the correlated energy chirp of the low charge electron bunch was reduced from approximately 330 keV/mm to zero. This result is in agreement with simulations. Calculations show that similar devices, properly scaled to account for the expected electron bunch charge and length, can be used to remove residual correlated energy spread at the end of the linacs used for free-electron lasers (FEL). Potentially, this technique could significantly simplify linac design and improve FEL performance.

Ultrashort x-ray pulses are a powerful tool for addressing grand challenges in science, e.g. control of materials and processes at the electron level, and the investigation and design of new forms of matter with tailored properties. Short (subpicosecond) pulses are central to many of the next generation light source initiatives (primarily free electron lasers (FEL)) that are based on linear accelerators. Subpicosecond bunches in FELs are produced by means of longitudinal bunch compression. Obtaining this compression requires the beam has to have a positive energy chirp: the longitudinal energy profile linearly correlated with the longitudinal coordinate with the head of the beam having a lower energy than the tail. This is realized by running the beam off-crest in the accelerating section. The residual chirp also helps to compensate for wakefield effects through the rest of the accelerating stage. Chirp is often measured in keV/mm, the correlated energy spread of the beam divided by its length. This small (but having a deleterious effect on the performance of the FEL) energy spread needs to be removed before the beam passes through the undulator. We consider here a concept for a passive wakefield device, the "wakefield silencer" [1] or "dechirper" [2], to perform this compensation. The device is to be installed, for example, at the output of the last compressor, or at the end of the linac. The silencer allows the removal of the remaining energy spread (~40 MeV/mm for FEL) after longitudinal bunch compression, using its self-wakefield - Cherenkov radiation generated by the bunch passing through a dielectric loaded [1,3] or corrugated wall [2] waveguide. We have demonstrated passive energy chirp correction by self-wakefield at the Brookhaven ATF facility [3]. In this paper we present the results of a *tunable* chirp correction experiment using a variable aperture planar dielectric loaded waveguide. The 330 keV/mm correlated energy spread of a 60 MeV, 54 pC beam at the BNL Accelerator Test Facility (ATF) was completely removed by adjusting the gap between the dielectric plates from 5.8 mm down to 1 mm.

While propagating along a dielectric-lined tube the ultra-relativistic electron bunch excites a set of waveguide modes with a phase velocity equal to electron velocity (the speed of light in the ultra-relativistic case). Unlike wakefield acceleration [4] and THz radiation generation [5] which employ primarily long-range wakefield radiation (rf energy left behind in the structure), here we exclusively consider short-range wakefields: the fields experienced by the beam itself. These short-range wakefields cause the head of the short beam to decelerate the tail. The wake amplitude ~ $\eta \cdot q/a^2$, where $\eta$ is the waveguide form factor, $q$ (C) is the charge and $a$ (m) is the characteristic size of the aperture. Hence tuning the transverse size of the aperture allows corresponding adjustment of the strength of the chirp corrector [6].

Removing a linear chirp requires that the decelerating field inside the bunch is also nearly linear along the bunch length. In Fig. 1(a), we present plots of the wakes (computed by the method of [7]) generated by a quasi-flat-top beam similar to that used in the design of the Next Generation Light Source (NGLS) [2] (approximately 150 μm flat-top core and 300 pC of charge) passing through the structure that we used in our experiment. Our dechirper was made of two 10 cm long alumina bars, 12 mm wide and 6.35 mm thick with metallized backs. The total gap size was remotely adjustable over the range 0-1 cm (Fig. 2(c)). The amplitude can be varied, with a negligible change in the profile of the decelerating field. The front slope of the decelerating field provides quasi-linear dechirping along the beam. For characterizing the dechirper we propose to use a figure of merit, the *dechirper strength*, measured in MeV/mm/m/nC. To obtain this unit, we start with the energy chirp, which is measured in MeV/mm (correlated energy variation per unit of length), normalized by the length of the dechirper in meters and then by the total charge of the beam in nC.

This is illustrated in Fig. 1(b), which shows a comparison of the wakes generated by the NGLS beam and a Gaussian beam with the same standard deviation, in the dechirper with a 1.9 mm gap. The difference in the decelerating field that a particle at $z_b$ experiences with respect to particle at $z_a$ is about 4 MV/m, while the longitudinal spacing between them is 0.17 mm. That means, for instance, that passing a 300 pC flat-top beam through a 1 m long structure would have removed a chirp of 4 MeV/0.17 mm. Therefore the chirp corrector strength is ~ 78.5 MeV/mm/m/nC. While this simple definition does not capture the effects of the current profile (in Fig. 1(b) the wake from a similar size Gaussian beam has a slightly different slope) it can be used as a figure of merit to describe the chirp corrector.

In the example in Fig. 1(a) the dechirper strength is 197 MeV/mm/m/nC for a 1 mm gap. In the case of a 2.8 mm gap the dechirper strength equals 40.7 MeV/mm/m/nC. In [2] a dechirper based on a corrugated wall structure was considered, assuming a larger (3 mm) aperture and the same 300 pC 150 μm flat-top NGLS beam. In this case a 6.65 m long structure is required to remove a 40 MeV/mm chirp, giving a dechirper strength of 20 MeV/mm/m/nC.

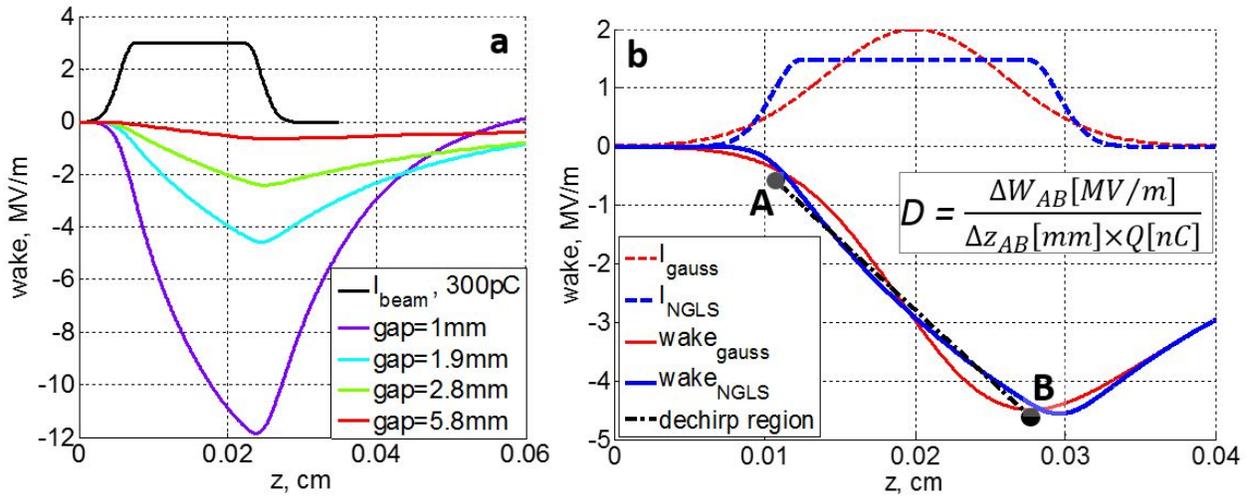

FIG. 1. (a) Short range wakefield as a function of the dechirper gap for the NGLS beam (300 pC, 150 μm quasi-flat-top). (b) dechirper strength calculation for the case of 1.9 mm gap and comparison of the wakes generated by the NGLS beam and Gaussian beam with the same standard deviation along z.

The structure has to be designed to yield a linear decelerating field for the specific shape of the bunch to be corrected. The principle of the chirp corrector is based on the action of each excited mode back on the beam itself. It is extremely important that the wavelength of excited mode is larger than the length of the beam to avoid field oscillations along the beam [3]. Consider a multimode waveguide loaded with a thick dielectric layer as the chirp compensation device. This structure would support a large number of modes with wavelengths larger that the bunch length (Fig.2(a)). The thick layer design is beneficial for small aperture operation, since it becomes practically impossible to host modes with large enough wavelengths in the case of a thin dielectric layer or corrugation.

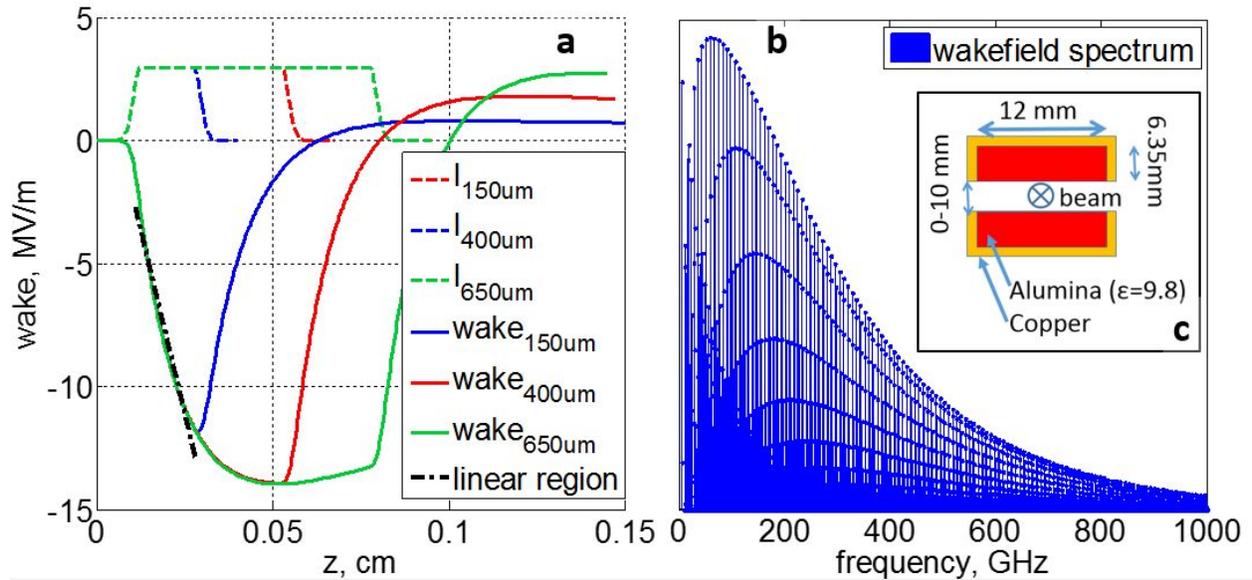

FIG. 2. (a) Wakes from flat-top beams of the same peak current, but different lengths (150, 400 and 650 μm). Long beams do not exhibit a linear decelerating wake. (b) Spectrum of modes excited by the beam in a dechirper with a 1 mm gap; (c) the transverse geometry of the dechirper.

For multimode structures there is a natural roll-off in the spectral content which is a consequence of the beam possessing a larger low frequency content compared to high frequency. For a successful dechirper one has to make sure that the modes with wavelengths longer than the beam dominate the short wavelength part of the spectrum. This is the case for our structure even for the ~ 550 μm long quasi-triangular beam used in the experiment (Fig. 1(b)). For a short triangular beam the self-decelerating field is always linear [3]. If we make the beam longer (Fig.

2(a)) the collective effect of modes with wavelengths smaller than the beam length becomes larger and leads to a non-linear decelerating field.

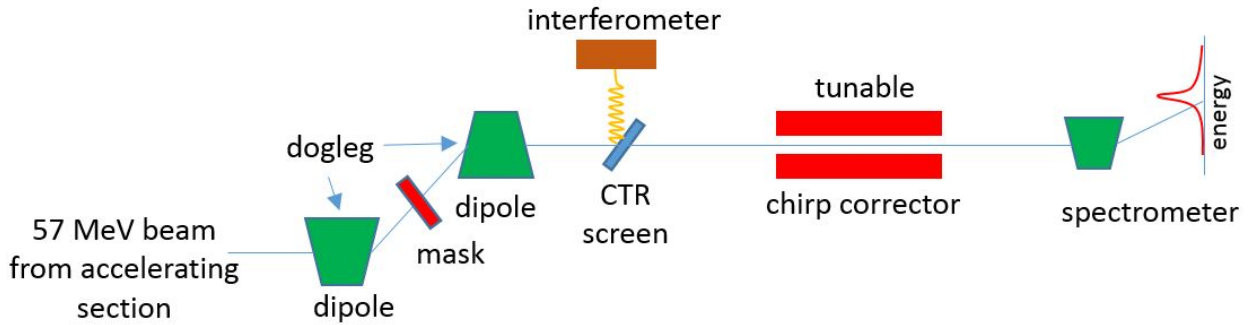

FIG. 3. Schematic of the experimental setup.

The experiment was performed at Accelerator Test Facility of Brookhaven. The general experimental layout is shown in Fig. 3. We used a 54 pC quasi-triangular shaped beam that was about 550 μm long. We accelerated the beam off-crest in the linac introducing a positive 330 keV/mm energy chirp. The current distribution is produced by passing the beam through the dogleg (two opposite sign dipole magnets) with a mask in between that removes the tails of the bunch. The beam size was calibrated by coherent transition radiation (CTR) interferometry. The beam was sent through the chirp corrector, a 10 cm long dielectric loaded waveguide with a variable gap size. The dielectric loading consisted of 12 mm wide, 6.35 mm thick alumina ($\varepsilon$ = 9.8) bars.

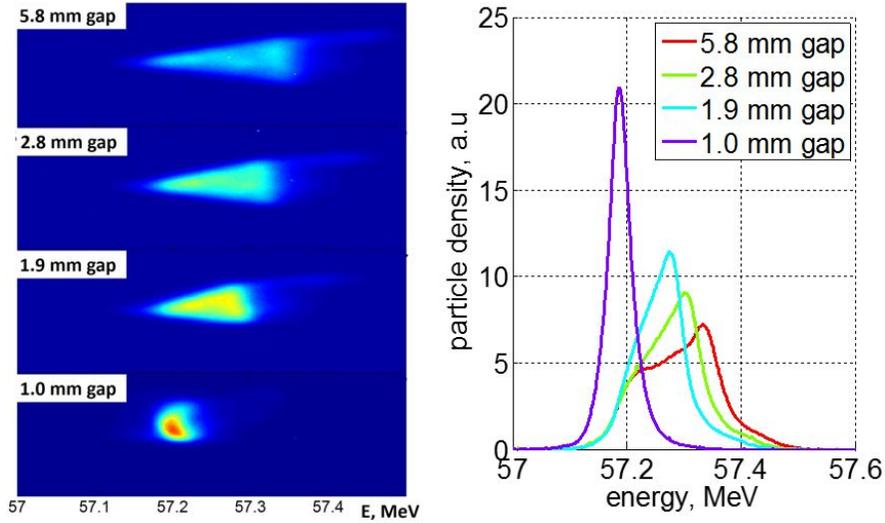

FIG. 4. (a) Energy spectra of the beam passing through a chirp corrector with various gap sizes: 5.8, 2.8, 1.9 and 1 mm. (b) Calibrated lineout of the beam spectrum. The energy spread is seen to be significantly reduced as the gap size approaches its optimum value.

We measured the energy spread using a magnetic spectrometer. It was observed that the energy spread decreased as we closed the chirp corrector gap from 5.8 mm to 1 mm. Fig. 3 shows our experimental data: the beam is detected by a phosphor screen and imaged by a camera (Fig.3(a)). On this image the horizontal dimension represents the particle energy, and the vertical dimension represents the vertical beam size. Due to the initial energy chirp one can see the correlation of the chirp value and the longitudinal projection of the beam on the spectrometer. When the dechirper gap was closed down to 1 mm we no longer observed any correlation and the energy spread was at a minimum. The energy chirp was calculated to be 330 keV/mm (165 keV over a 500 micron long triangular beam). This 330 keV/mm was compensated in the 10 cm long dechirper. Since the total beam charge was 54 pC, the dechirper strength in this case is 61±6 MeV/mm/m/nC.

There is a simple theoretical model that fits well the experimental data presented above. The beam is modeled by taking the measured current shape (Fig. 5(a)) and applying a linear energy chirp to it along with an estimated uncorrelated energy spread of 30 keV. Then we use wakes simulated for the measured current shape and dechirp the constructed beam numerically using those wakes (Fig.5(b)). The main source of inaccuracy arises from the bunch shape measurement, estimated as ~10%. This model fits the data quite well within the error bars (Fig. 5(b)). We plot the standard deviation of the beam as a function of the dechirper gap for both the experimental and numerical data (Fig.5(b)).

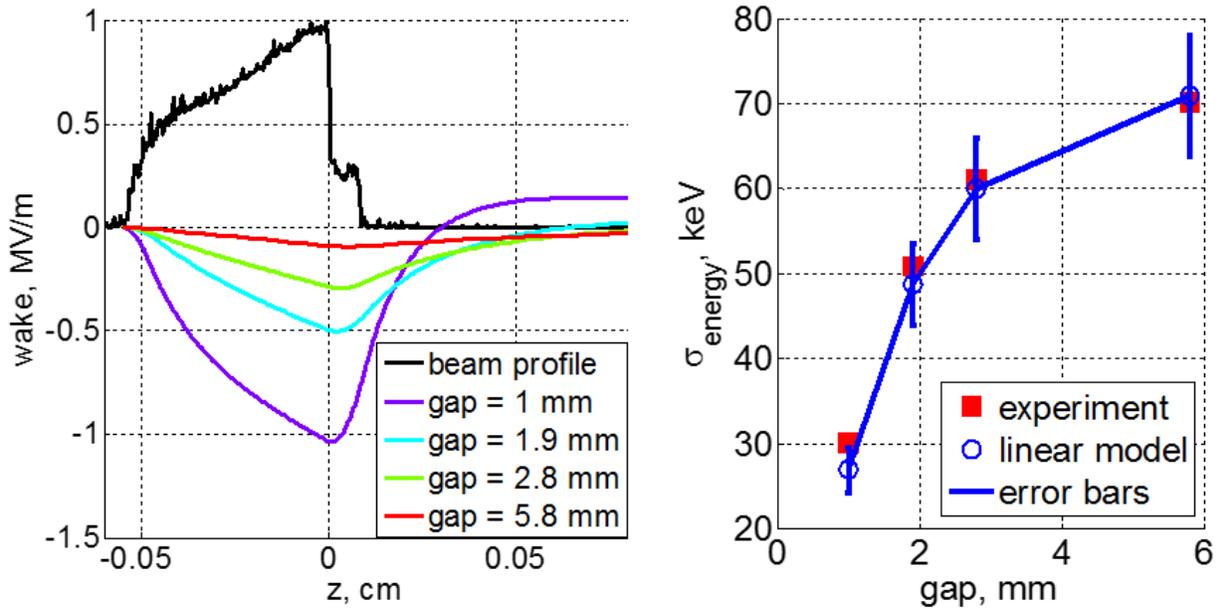

FIG. 5. (a) Wake computed from the measured experimental current profile. (b) Energy spread as a function of the gap size; comparison of experimental results and the linear theory.

To summarize, we have successfully demonstrated a technique for tunable chirp correction. A beam with 330 keV/mm positive energy chirp traversed a dielectric loaded structure with a variable aperture. By gradually changing the transverse gap size of the dechirper from 5.8 mm down to 1.0 mm we reduced the correlated energy spread from 330 keV/mm down to 0. The residual energy spread corresponded to the estimated 30 keV thermal energy spread of the beam.

Acknowledgements: Euclid Techlabs LLC acknowledges support from DOE SBIR program grant #DE-SC0006299 and U.S. Department of Energy Office of Science under Contract No. DE-AC02-06CH11357.